\documentclass[12pt]{article}
\usepackage{amsmath}
\usepackage{latexsym}
\usepackage{epsf}
\usepackage{amsfonts}
\begin{document}
\vspace *{2cm} 
\begin{flushright}
UTEXAS-HEP-99-9
\end{flushright}
\begin{center}
{\Large \bf Einstein-Brans-Dicke Cosmology}

\bigskip

Yungui Gong \footnote{Email: ygong@physics.utexas.edu} 

Physics Department, University of Texas at Austin, Austin, TX 78712, U.S.A.
 
\bigskip
 
\vskip 0.5in
{\Large\bf Abstract}
\end{center}
 
\parindent=4ex

We studied the Einstein-Brans-Dicke cosmology in detail. The difference
of the evolution of the universe is
significant between Einstein-Brans-Dicke cosmology and standard big-bang
model during the radiation-dominated era. The power-law evolution of the
scale factor is fast enough to solve the cosmological puzzles and slow
enough to avoid the graceful exit problem. However, the constraints from
the satisfactory bubble distribution ($\beta^2>0.25$) and the solar
system observations ($\beta^2<0.002$) are mutually exclusive. This
suggests that this kind of inflationary model is ruled out. We also
clarify the distinction between Einstein frame and Jordan frame in
Brans-Dicke theory.

\vskip 6cm

\pagebreak
 
In Brans-Dicke (BD) gravity, the effective
gravitational constant $G_{eff}=\phi ^{-2}$,
varies as the BD field $\phi$ evolves. The evolution of the
scalar field slows down the expansion
rate of the universe during inflation, and allows 
nucleation of bubbles to end the inflationary era. But it was soon
found that the bubbles could lead to unacceptable distortions of
the microwave background \cite{d10}.
A large number of inflationary models were proposed in the framework
of multi-scalar tensor gravity to solve the problem \cite{d41}-\cite{d9}.
For instance, the introduction of a potential for the scalar
field $\phi$ and a scalar field dependent coupling constant $\omega(\phi)$
solved some problems. 
In \cite{d9}, D. La considered the BD cosmology in Einstein frame,
but he didn't analyze the constraints from cosmological models
in Einstein frame
\footnote{Einstein frame is also called Pauli frame in the literature}.
Instead
he used the constraints from the original Jordan-Brans-Dicke (JBD) inflation.
In the literature, some people considered
the inflationary models in Einstein frame in order to solve equations
easily, but they analyzed their final results in Jordan frame 
because most people insist that the Jordan frame is the physical frame
to keep the equivalence principle. In fact, the equivalence principle
can be kept in Einstein frame if we use Einstein frame as the physical
frame.
Sometimes people just mixed up Jordan frame and Einstein frame.
As showed by Cho and Damour etc., Pauli metric represented the
spin-two massless graviton \cite{d4}\cite{dn}. They also showed that the two
frames were not conformal invariant for the case used in inflationary models.
For arguments in favor of Einstein frame as the 
physical frame, see \cite{frame}.
In Kaluza-Klein unification, one must identify the
physical 4-dimensional metric as Pauli metric $g_{\mu\nu}$ in
order to describe Einstein gravity \cite{d35}-\cite{d14}. 
Apart from the higher dimensional
Kaluza-Klein theory, the Einstein-Brans-Dicke (EBD)
like theory may also derived
from the induced gravity and $R^2$ gravitational theories.
In this paper, we choose Einstein frame as the physical one and
refer the cosmology based on the EBD gravity
as EBD cosmology. We analyze the detailed evolutions of the universe
during the radiation-dominated (RD), matter-dominated (MD) and 
the inflationary epochs. 
The physical differences from
these two frames were discussed in \cite{d4}\cite{frame}\cite{d1}\cite{yzg}. 
The distinctive features of the EBD gravity are:
(1) a massive test particle deviates from geodesic motion,
and a photon follows geodesic motion. 
Cho and Magnano and Sokolowski derived that a photon remained geodesic
motion by using
the equation of motion for the BD scalar field \cite{frame}\cite{d36}. 
The derivation is wrong because
the use of the equation of motion for the BD scalar field means that
the test particle is the source of spin-0 gravitational field. In fact,
we should consider the motion of a test particle in some known background,
not the interaction between the test particle and the background space-time.
The geodesic motion of a photon is the consequence of $ds^2=0$ and
the conformal invariance of $ds^2=0$. In \cite{yzg}, we got the wrong
results about the deflection angle and the time delay of radar echo.
They should be different from those in general relativity by a factor
$1-\beta^2/2$. Therefore, the constraint on $\beta^2$ from both experiments 
is $\beta^2\leq 0.002$.
(2) The coupling constant $\omega$ in Brans-Dicke theory must be
positive, but in EBD theory, Pauli metric can be
defined even when $\omega$ is negative but larger than
$-3/2$. (3) the dilaton $\sigma$ does appear in the matter Lagrangian.
(4) the dilaton field no longer plays the role of time-varying
gravitational coupling constant.

The JBD Lagrangian is given by
\begin{equation}
\label{bdlagr}
{\cal L}_{BD}=-\sqrt{-\gamma}\left[\phi{\tilde {\mathcal R}}+
\omega\,\gamma^{\mu\nu}
{\partial_\mu\phi\partial_\nu\phi\over \phi}\right]-{\cal L}_m(\psi,\,
\gamma_{\mu\nu}).
\end{equation}
The above Lagrangian (\ref{bdlagr}) is conformal invariant under
the conformal transformations,
$$g_{\mu\nu}=\Omega^2\gamma_{\mu\nu},\quad \Omega=\phi^\lambda,~~
(\lambda\neq {1\over 2}),\quad \sigma=\phi^{1-2\lambda},
\quad {\bar \omega}={\omega-6\lambda(\lambda-1)\over (2\lambda-1)^2}.$$
For the case $\lambda=1/2$, we make the following transformations
\begin{subequations}
\begin{gather}
\label{conformala}
g_{\mu\nu}=e^{a\sigma}\gamma_{\mu\nu},\\
\label{conformalb}
\phi={1\over 2\kappa^2}e^{a\sigma},
\end{gather}
\end{subequations}
where $\kappa^2=8\pi G$, $a=\beta\kappa$, and $\beta^2=2/(2\omega+3)$. 
Remember that the JBD Lagrangian is not
invariant under the above transformations (\ref{conformala})
and (\ref{conformalb}). After the conformal transformations
(\ref{conformala}) and (\ref{conformalb}),
we get the EBD Lagrangian 
\begin{equation}
\label{1}
{\cal L}= \sqrt{-g} \left[-\frac{1}{2\kappa^2}{\mathcal R}
-\frac{1}{2}g^{\mu\nu}\partial_\mu \sigma \partial_\nu \sigma\right]
-{\cal L}_{m}(\psi, e^{-a\sigma}g_{\mu\nu}).
\end{equation}
In this frame, $h_{\mu\nu}=g_{\mu\nu}-\eta_{\mu\nu}$
represents the spin-2 massless graviton. That's one of the reasons why
we use the Einstein frame as the physical frame.
In general, we may use the matter Lagrangian 
${\cal L}_{m}(\psi, g_{\mu\nu})$ to keep the equivalence principle if we
identify the metric $g_{\mu\nu}$ as the physical one. However, we
will lose the interactions between the dilaton field $\sigma$ and the matter
fields $\psi$. In order to avoid this and see the differences between
the two frames, I use the Lagrangian (\ref{1})
as the basis of EBD cosmology.
In this paper, I consider the simplest case, i.e., a constant coupling
constant without potential for the dilaton field $\sigma$. The generalization
to a more complicated multi-scalar tensor gravity in Jordan
frame can be found in \cite{dn}. For the cosmological models in the
context of general scalar-tensor
theory in Jordan frame, see \cite{jdb}.

Based upon the homogeneous and isotropic
Friedman-Robertson-Walker spacetime
\begin{equation}
\label{rwcosm}
ds^2=-dt^2+R^2(t)\left[{dr^2\over 1-k\,r^2}+r^2\,d\Omega\right],
\end{equation}
and the perfect fluid
$T_m^{\mu\nu}=e^{-2\,a\sigma}[(\rho+ p)\,U^\mu\,U^\nu + p\,g^{\mu\nu}]$
as the matter source,
we can get the evolution equations
of the universe from the action (\ref{1})
\begin{gather}
\label{3.1a}
H^2+{k \over R^2}={\kappa^2 \over 3}\left({\frac 1 2}\dot{\sigma}^2
+e^{-2a\sigma}\rho\right),\\
\label{3.1b}
\ddot{\sigma}+3H\dot{\sigma}={\frac 1  2}a e^{-2a\sigma}(\rho-3p),
\end{gather}
where $\rho$ is the 
mass-energy density and $p$ is the pressure. 
The motion of the matter field satisfying the
covariant conservation law 
$$\nabla_\nu\left[T _{\sigma}^{ \mu \nu} +T _m^{\mu \nu}
\right]=0$$
with
$$\nabla_\nu T_{\sigma}^{ \mu \nu}=g^{\mu\nu}\partial_\nu\sigma\Box\sigma,$$
becomes
\begin{equation}
\label{3.1c}
\dot{\rho} +3H(\rho + p)={\frac 3 2}a\dot{\sigma}(\rho +p).
\end{equation}
If we are given a state equation for the matter $p=\gamma\rho$, then the
above equation gives us a first integral,
\begin{equation}
\label{3.1ci}
\rho\,R^{3(\gamma+1)}\,e^{-3a(\gamma+1)\sigma/2}=C_2.
\end{equation}
Combining Eqs. (\ref{3.1a}), (\ref{3.1b}) and (\ref{3.1c}), we get another first
integral for the flat universe $k=0$,
\begin{equation}
\label{3.1i}
R\,e^{-a(1-\gamma)\sigma/\beta^2(1-3\gamma)}=C,
\end{equation}
where the above equation is valid for $-1\le \gamma <1-2/(3+\sqrt{6}/\beta)$ 
and $\gamma\neq 1/3$.

From Eq. (\ref{3.1a}), we have
\begin{equation}
\label{3.0a}
{k\over H^2 R^2}={8\pi G\over 3H^2}({\frac 1 2}\dot{\sigma}^2
+e^{-2a\sigma}\rho)-1\equiv \Omega -1,
\end{equation}
where $\Omega\equiv 8\pi G({\frac 1 2}\dot{\sigma}^2 +e^{-2a\sigma}\rho)/
3H^2=({\frac 1 2}\dot{\sigma}^2+e^{-2a\sigma}\rho)/\rho_{c}.$
From the above expressions, we see how the $\sigma$ field
contributes to the matter source. The EBD cosmology was discussed in 
\cite{gong}, here we give more detailed analyses and make some corrections.

\vskip .2in

\parindent=0ex

\section{Matter-Dominated Epoch}

\parindent=4ex

In MD epoch, we have the state equation $p=0$.
The solutions to Eqs. (\ref{3.1a})-(\ref{3.1c}) for flat universe $k=0$ are
\begin{gather}
\label{3.3h}
\rho(t)=\rho_p\left[1+{6+\beta^2\over 4}\,H_p(t-t_p)\right]^{-6(2-\beta^2)
/(6+\beta^2)},\\
\label{3.3i}
R(t)=R_p\left[1+{6+\beta^2\over 4}\,H_p(t-t_p)\right]^{4/(6+\beta^2)},\\
\label{3.3j}
e^{a\sigma}=(16\pi)\left[1+{6+\beta^2\over 4}\,H_p(t-t_p)\right]^
{4\,\beta^2/(6+\beta^2)},\\
\label{mdht}
H(t)={H_p\over 1+{6+\beta^2\over 4}\,H_p(t-t_p)}.
\end{gather}
where $R_p$, $H_p$ and $t_p$ are 
the present radius of the universe, the present
Hubble constant and the present age of the universe, respectively, and
$e^{a\,\sigma_p}=16\pi$. If we let $t_p=4H_p^{-1}/(6+\beta^2)$, then we have
\begin{equation}
\label{mdsolu}
\rho(t)=\rho_p\left({t\over t_p}\right)^{-6(2-\beta^2)/(6+\beta^2)},~
R(t)=R_p\left({t\over t_p}\right)^{4/(6+\beta^2)},~
e^{a\sigma}=(16\pi)\left({t\over t_p}\right)^{4\,\beta^2/(6+\beta^2)}.
\end{equation}
It is obvious that the present age of the universe given by EBD theory
is a little less than that given by the standard big-bang model.
Because the smallness of the value of $\beta^2$ determined by 
the present experiments, we obtain the approximate solutions
\begin{gather}
\label{3.3l}
\rho\approx\rho_{p}[1+{\frac 3 2}H_p (t-t_p)]^{-2}=\rho_{p}
\left({t\over t_p}\right)^{-2},\\
\label{3.3m}
R\approx R_p[1+{\frac 3 2}H_p (t-t_p)]^{2/3}
=R_p\left({t\over t_p}\right)^{2/3},\\
\label{3.3n}
\sigma\approx {\ln{(16\pi)}\over \beta\,\kappa} +{2\beta\over 3\kappa}
\ln{[1+{\frac 3 2}H_p (t-t_p)]}\approx {\ln{(16\pi)}\over a}.
\end{gather}

From the above results, we see
that the contribution of dilaton field is negligible and the
evolution of the universe is almost indistinguishable from
the usual hot-big-bang model based on Einstein gravity during MD epoch.

\vskip 0.2in
\parindent=0ex

\section{Radiation-Dominated Epoch}

\parindent=4ex

For the RD epoch, the equation of state for radiation is
$\rho=3p.$ After using the equation of state, we get solutions to 
Eqs. (\ref{3.1b}) and (\ref{3.1c}) for $k=0$
\begin{gather}
\label{3.01}
R(t)^3\dot{\sigma}(t)=C_1,\\
\label{3.02}
\rho(t)\,R^4(t)\,e^{-2a\sigma}=C_2,
\end{gather}
where $C_1$ and $C_2>0$ are the integration constants.
$C_1$ can be positive, negative or zero determined by
the initial condition of the universe.
If we choose $C_1=0$, then we find that $\sigma$ is a
constant, so the solutions are the same as those of
the standard big-bang model during RD era.
Combining Eqs. (\ref{3.01}), (\ref{3.02}) and (\ref{3.1a}), we
get 
\begin{equation}
\label{rdradius}
\left({\dot{R}\over R}\right)^2={\kappa^2\over 3}\left({C_1^2\over 2\,R^6}+
{C_2\over R^4}\right).
\end{equation}

\begin{figure}[htb]
\vspace{-0.3in}
\begin{center}
$\begin{array}{cc}
\epsfxsize=2.7in \epsffile{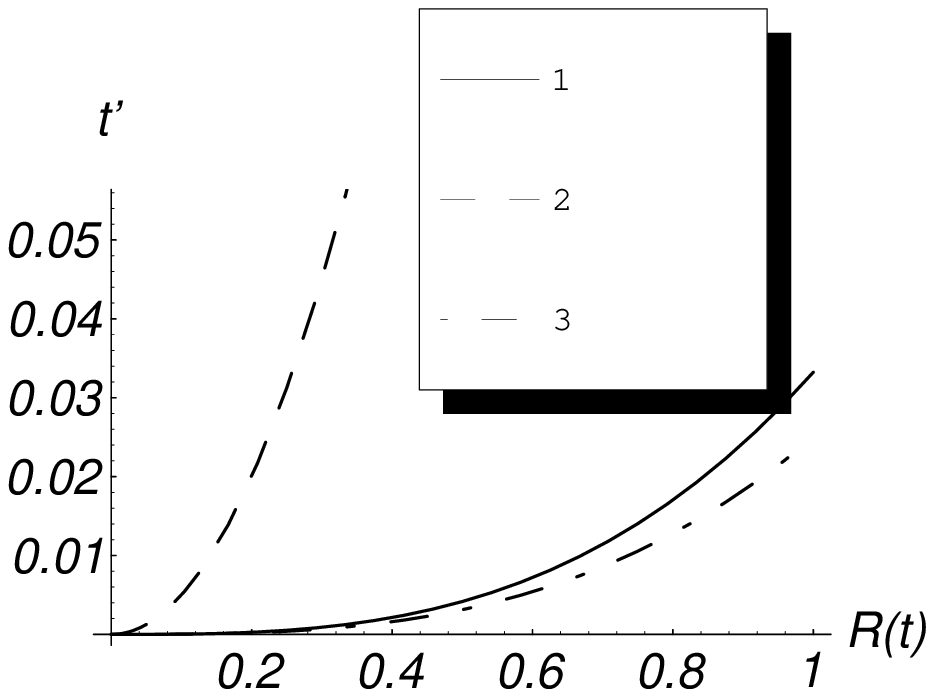}&
\epsfxsize=2.7in \epsffile{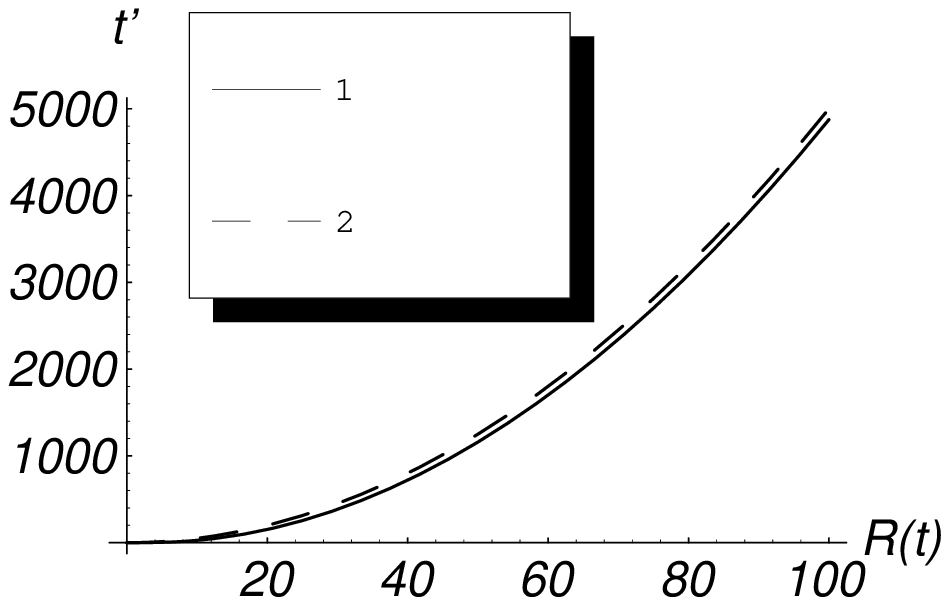}
\end{array}$
\end{center}
\vspace{-0.5in}
\caption{The time ${\bar t}$ as the function of the scale factor $R$.
Curve 1 refers to $\alpha=10$, curve 2 refers to $\alpha=0$ and the third
curve in the left figure is $R^3/4\alpha$ with $\alpha=10$.}
\label{trfig}
\end{figure}

\begin{figure}[htb]
\vspace{-0.5in}
\begin{center}
$\begin{array}{cc}
\epsfxsize=2.7in \epsffile{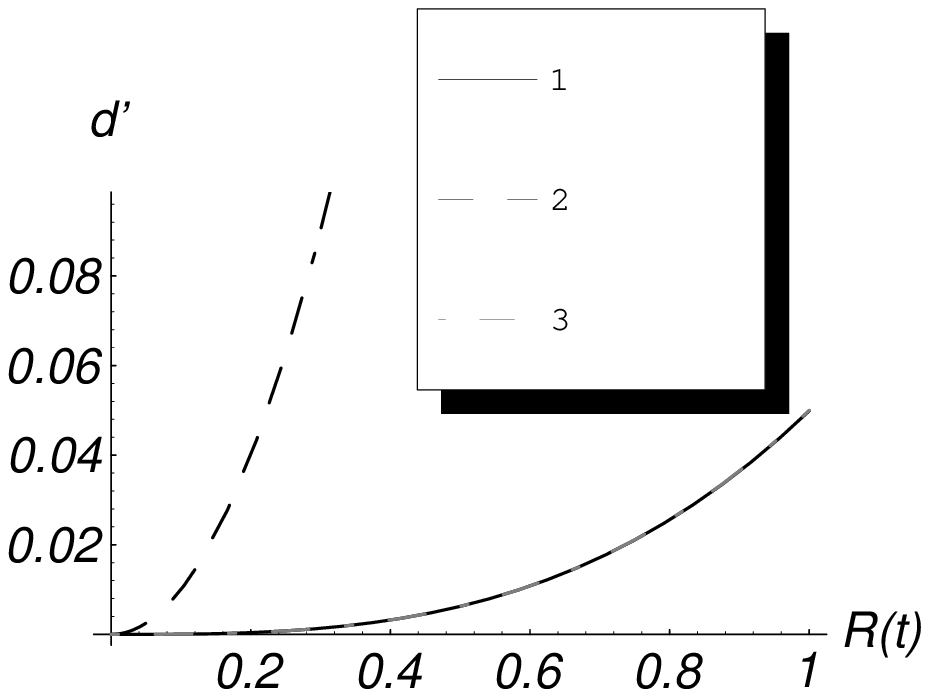}&
\epsfxsize=2.7in \epsffile{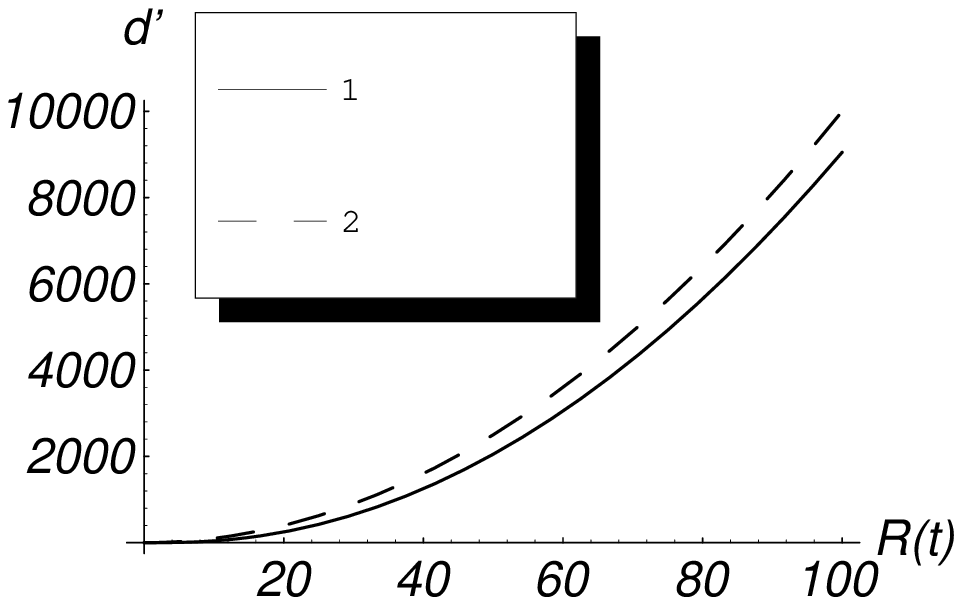}
\end{array}$
\end{center}
\vspace{-0.5in}
\caption{The horizon ${\bar d}_H$ as the function of the scale factor $R$.
Curve 1 refers to $\alpha=10$, curve 2 refers to $\alpha=0$ and the third
curve in the left figure is $R^3/2\alpha$ with $\alpha=10$.}
\label{drfig}
\end{figure}

It is easy to solve the above equation if we use the cosmic time
defined by $dt=R(\eta)d\eta$. The solutions to the above equations
(\ref{3.01})-(\ref{rdradius}) in
terms of the cosmic time $\eta$ are
\begin{gather}
\label{rdscale}
R(\eta)=\left[{C_2\over 3}\kappa^2\eta^2+
{2\over \sqrt{6}}|C_1|\kappa\eta\right]^{1/2},\\
\label{rdscalar}
\sigma=\sigma_0\pm{3\over \sqrt{6}\kappa}
\ln{\eta\over \eta+{\sqrt{6}|C_1|\over C_2\kappa}},\\
\label{rddensity}
\rho(\eta)={C_2\,e^{2a\sigma}\over R^4(\eta)},
\end{gather}
\begin{gather}
\begin{split}
t&={1\over 2C_2}\sqrt{{3\over C_2\kappa^2}}\bigg[
({1\over \sqrt{6}}C_2\kappa\eta+{|C_1|\over 2})\sqrt{
{2\over 3}C_2^2\kappa^2\eta^2+{4\over \sqrt{6}}|C_1|C_2\kappa\eta}
\\
&-{C_1^2\over 2}\ln\left({2\over \sqrt{6}}C_2\kappa\eta+|C_1|
+\sqrt{
{2\over 3}C_2^2\kappa^2\eta^2+
{4\over \sqrt{6}}|C_1|C_2\kappa\eta}\right)\bigg]
+{C_1^2\over 4C_2}\sqrt{{3\over C_2\kappa^2}}\ln|C_1|,
\end{split}\\
\label{rdrt}
R(t)\sqrt{R^2(t)+\alpha^2}-\alpha^2\ln(R(t)+\sqrt{R^2(t)+\alpha^2})+
{\alpha^2\over 2}\ln\,\alpha^2
=2\sqrt{{C_2\over 3}}\kappa\,t=2\,{\bar t}
\end{gather}
where ${\bar t}=\sqrt{C_2}\,\kappa t/\sqrt{3}$, $\alpha^2=C_1^2/(2\,C_2)$,
$\sigma_0$ is another integration constant determined by the
initial condition of the dilaton field $\sigma$ and the sign in the equation
(\ref{rdscalar}) is the same as the sign of 
$C_1$.
From Eq. (\ref{rdrt}), it is clear that (see Fig. \ref{trfig})
$$2\,{\bar t}
\approx 
\begin{cases}
{\displaystyle R^3(t)\over \displaystyle 2\alpha}\ll R^2(t)&
{\rm if}\ R(t)\ll \alpha,\\
R^2(t) &{\rm if}\ R(t)\gg \alpha.
\end{cases}$$
In the early times, (here we suppose $C_1$ is not very large
so that $R(t)$ will exceed $\alpha$ in a very short time), the universe
in EBD cosmology
evolves much faster than that in standard cosmology.
After some time ($R(t)\gg\alpha$), 
the solution will evolve to
the solution with $C_1=0$ asymptotically, i.e., we have the same evolution
as that given by the standard model. From Eq. (\ref{rdradius}) and
(\ref{rdrt}), we can get the proper distance to the horizon measured at
time $t$
\begin{equation}
\label{rdhorizon}
d_H(R)=R(t)\int_{0}^{t}{dt'\over R(t')}=
{\sqrt{3}\over\sqrt{C_2}\kappa}R\,(\sqrt{R^2+\alpha^2}-\alpha).
\end{equation}
Let ${\bar d}_H(R)=\sqrt{C_2}\,\kappa d_H(R)/\sqrt{3}$,
we have ${\bar d}_H(R)\approx  R^3/2\alpha\ll R^2$ if $R\ll\alpha$.
Therefore, the horizon distance in the early times
is much smaller than that in standard
model, we will need more e-foldings to solve the horizon problem.
If $R\gg\alpha$, ${\bar d}_H(R)\approx R^2$ (see Fig. \ref{drfig}).
From the above analyses,
we know that during the early times, the expansion of
the universe is faster than that
in standard big bang model. This means that at the same temperature
or same size, the EBD universe is younger.

\vskip .2in
\parindent=0ex

\section{Inflationary Epoch}

\parindent=4ex

As the universe enters the inflationary epoch, the energy 
density approaches the false-vacuum energy density $\rho_f=-p_f$=
const. The solutions to Eqs. (\ref{3.1a})-(\ref{3.1c}) for flat universe
are
$$e^{a\sigma}=2\kappa^2\,{\tilde m}^2_{Pl}(1+2\,\beta^2\,H_B\,t),$$
or
\begin{gather}
\label{3.2c}
\sigma=\sigma_B +{1\over \beta\kappa}\ln{(1 +2\,\beta^2\,H_B\,t)},\\
\label{3.2d}
R(t)=R(B)(1+2\,\beta^2\,H_B\,t)^{\displaystyle 1\over \displaystyle
2\,\beta^2},
\end{gather}
where $H_B=\sqrt{\rho_f}/(2\,\kappa\,{\tilde m}^2_{Pl}\sqrt{3-2\beta^2})$ 
is the Hubble
parameter at the beginning of inflation, $t=0$ (here I set the beginning
of the inflation to be time scale zero), ${\tilde m}_{Pl}$ (greater
than the scale of phase transition) is an arbitrary
integration constant corresponding to the effective Planck mass
at the beginning of inflation, and $\sigma_B=\ln{(2\,\kappa^2
\,{\tilde m}^2_{Pl})}/a$
is the value of dilaton field at the beginning of inflation.
Because the variation of the dilaton field $\sigma$ is very small during
the MD era and the late times of the RD epoch,
we may suppose that at the end of inflation, the dilaton field becomes
$\sigma(t_e)\approx \ln(16\pi)/a$. If $C_1$ is large enough that the universe
will stay at the range $R(t)<\alpha$ during most of the times of RD era,
then the above assumption is not true.

For short times, $t<t_c\equiv 1/2\,\beta^2\,H_B$, 
$R(t)\sim\exp{(H_B t)}$, which is the Einstein-de-Sitter
inflation. However, for $t>t_c$, $R(t)$ crosses over to power-law 
expansion, $R(t)\sim (t/t_c)^{{1 \over 2\,\beta^2}}.$
The cross-over from exponential to power-law expansion changes the
rate at which bubble nucleation converts the universe from false- 
to true- vacuum phase. The phase transition will be completed when 
the nucleation rate per Hubble volume per Hubble time
\begin{equation}
\label{nucl}
\epsilon(t)={\lambda_0\over H(t)^4}\sim \lambda_0(2\beta^2)^4\,t^4\sim 1.
\end{equation}
Let us suppose that the phase transition ends at time
(here we follow Weinberg's method \cite{d10})
\begin{equation}
\label{endtime}
t_e=q\,\lambda_0^{-1/4}(2\beta^2)^{-1},
\end{equation}
where $q$ is an order of unity constant and $\lambda_0$ is the nucleation
rate per unit volume per unit time.
The consistency condition by the dilaton field at the
end of inflation gives
\begin{gather}
\label{dilaton}
\sigma(t_e)=\sigma_B+{1\over a}\ln(1+2\beta^2\,H_B\,t_e)\approx
{1\over a}\ln(16\pi),\\
\label{htrel}
2\beta^2\,H_B\,t_e\approx {1\over G\,{\tilde m}_{Pl}^2}\gg 1.
\end{gather}
Combining Eqs. (\ref{endtime}), (\ref{htrel}) and the definition
of $H_B$, we get
\begin{equation}
\label{lamd}
\lambda_0=\left({1\over 96\pi}\right)^2\,q^4\,\rho_f^2\,G^2.
\end{equation}
During the phase transition, the scale factor will increase by a factor
\begin{equation}
\label{vdscale}
{R(t_e)\over R(0)}=(1+2\beta^2\,H_B\,t_e)^{1/2\beta^2}\approx
(2\beta^2\,H_B\,t_e)^{1/2\beta^2}.
\end{equation}
The requirement that the scale factor increases at least by a factor
of 65-e foldings gives us the constraint
\begin{equation}
\label{beta1}
\beta^2 <{1\over 65}\ln\left({T_{Pl}\over T_c}\right)=0.14,
\end{equation}
where $T_{Pl}=10^{19}$ Gev is the Planck energy and $T_c=10^{15}$ Gev
is the energy scale for GUT phase transition.
The probability of a point remaining in the
false-vacuum phase during a bubble nucleation process beginning at
time $t_B$ is
\begin{equation}
\label{3.37}
p(t)=\exp\left[-\int_{t_B}^t dt' \lambda(t')R^3(t'){4\pi \over 3}\left[
\int_{t'}^t {dt''\over R(t'')}\right]^3\right],
\end{equation}
where $\lambda(t)$ is the nucleation rate per unit time per unit
volume, approximately constant ($\sim\lambda_0$) during the inflationary phase.
Combining Eqs. (\ref{3.2d}), (\ref{3.37}) and (\ref{lamd}), we get
\begin{equation}
\label{3.38}
p(t)=\exp\left[-{\pi \over 3}{\delta\over 2 \beta^2}\left(y^4\,g(\beta)-1
+O[(1+2\beta^2\,H_B\,t)^{1-1/2\beta^2}]\right)\right],
\end{equation}
where $y=1+2\,\beta^2\,H_B t$, 
$\delta=(q/(2\,\beta^2\,H_B\,t_e))^4$, and 
$$g(\beta)=1-{24\beta^2\over 6\beta^2+1}+{12\beta^2\over 1+\beta^2}-
{8\beta^2\over 3+2\beta^2}.$$ 
If $2\beta^2\,H_B\,t$ is large
and $\beta$ is small, we have
\begin{equation}
\label{prob}
p(t)\approx \exp\left[-{\pi \over 3}{q^4\over 2\beta^2}\left(
{t\over t_e}\right)^4\right].
\end{equation}

From the bounds on the anisotropy of the microwave background,
we will get a constraint if we require that no more than $10^{-5}$
of space was still undergoing thermalization at the recombination
$T\approx 4000$ K,
\begin{equation}
\label{beta2}
\beta^2>{1\over 2+{8\over 5}\log_{10}(T_c/T)}\approx 0.025\quad
{\rm or}\ \omega<{1\over 2}+{8\over 5}\log_{10}(T_c/T).
\end{equation}	

Remember that the solar system observation requires
$\beta^2<0.002$, so the EBD inflation can't 
avoid the big-bubble distribution problem either.

It is true that we can get the Eqs. (\ref{3.1a})-(\ref{3.1c}) and all the
solutions in this paper from the corresponding equations and solutions
in the original JBD cosmology by the transformations
\begin{equation}
\label{jtog}
dt=e^{a\sigma/2}d{\tilde t},\qquad R(t)=e^{a\sigma/2}{\tilde R}[{\tilde t}(t)],
\end{equation}
and the transformation (\ref{conformalb}),
where ${\tilde R}({\tilde t})$ is the scale factor in JBD cosmology. 
That is, our solutions in terms
of $t$ can be derived from the solutions in terms of ${\tilde t}$ in original
JBD cosmology by the above transformations. This is
easily understood. Note that in this paper
we use the Robertson-Walker metric (\ref{rwcosm})
and our Lagrangian is related to the JBD Lagrangian
by the transformations (\ref{conformala}) and (\ref{conformalb}).
Under the transformations (\ref{conformala}) and (\ref{conformalb}),
the Robertson-Walker metric becomes
$$ds^2=e^{a\sigma}d{\tilde s}^2=
e^{a\sigma}\left\{-d{\tilde t}^2+{\tilde R}^2({\tilde t})\left[{dr^2\over 1-k\,r^2}+
r^2\,d\Omega\right]\right\}.$$
Therefore, if we make the transformations (\ref{jtog}) 
and (\ref{conformalb}), we can get
the solutions to EBD cosmology from the corresponding solutions to
the JBD cosmology.
So the two cosmological solutions are related by the transformations
(\ref{jtog}) and (\ref{conformalb}). 
Although we have the relationships (\ref{jtog}) and (\ref{conformalb})
between the two cosmological models, the actual physical contents are different if we
use different identifications of the physical metrics. 
This point is obvious
from the results in this paper.
The comoving coordinates
are (${\tilde t}$, $r$, $\theta$, $\phi$) in JBD cosmology and the comoving
coordinates are ($t$, $r$, $\theta$, $\phi$) in EBD cosmology.
Note that the coordinate symmetry is broken in cosmology.
As showed in this paper, 
the problem arising from bubble
distribution is also unavoidable in EBD inflation. 

In EBD cosmology, I find that the expansion of the universe is
faster than that given by standard cosmology during the early times of RD
era. This has an important dynamical effect upon the early universe.
The total entropy in EBD cosmology defined as $S=e^{-2a\sigma}(\rho+p)R^3/T$
is conserved. Since we have $e^{-2a\sigma}\rho\sim T^4$, so 
$R(t)\sim T^{-1}(t)$. 
The faster expansion makes $T(t)$ decrease, i.e., a given temperature
will occur at an earlier epoch. 

\bigskip

{\noindent \large \bf Acknowledgement}

\smallskip

I would like to thank Prof. Ne'eman for his useful comments on this paper.

\end{document}